\documentclass[a4paper]{article}
\usepackage{graphicx}
\usepackage{amsmath}

\begin{document}

\title{Relative entropy of entanglement of a kind of two qubit entangled states }
\author{Xiao-yu Chen$^1$, Li-min Meng$^2$,Li-zhen Jiang$^1$,Xiang-jun Li$^1$ \\
 \small 1) School of Science, China Institute of Metrology, Hangzhou,
310018,China;\\
\small 2)College of Information Engineering, Zhejiang University
of Technology, Hangzhou 310032, China;}
\date{}
\maketitle

\begin{abstract}
We in this paper strictly prove that some block diagonalizable two qubit
entangled state with six none zero elements reaches its quantum relative
entropy entanglement by the a separable state having the same matrix
structure. The entangled state comprises local filtering result state as a
special case.


\end{abstract}

\section{Introduction}

Quantum relative entropy function has many applications in the problems of
classical and quantum information transfer and quantum data compression \cite
{Lin}. The relative entropy has a natural interpretation in terms of the
statistical distinguishability of quantum states; closely related to this is
the picture of relative entropy as a distance measure between density
operators. Based on the relative entropy, a nature measure of entanglement
called the relative entropy of entanglement was proposed. This entanglement
measure is intimately related to the entanglement of distillation by
providing an upper bound for it. \ It tells us that the amount of
entanglement in the state with its distance from the disentangled set of
states. In statistical terms, the more entangled a state is the more it is
distinguishable from a disentangled state\cite{Vedral}. However, except for
some special situations\cite{Wu}, such an entanglement measure is usually
very difficult to be calculated for mixed state. The relative entropy of
entanglement is defined as \cite{Vedral1}
\begin{equation}
E_r\left( \rho \right) =\min_{\sigma \in \mathcal{D}}S\left( \rho \left\|
\sigma \right. \right) =\min_{\sigma \in \mathcal{D}}Tr\rho (\log \rho -\log
\sigma ).,
\end{equation}
where the minimization is performed over the set $\mathcal{D}$ of separable
states ($\log $ denotes the natural logarithm throughtout this paper). When
any separable state $\sigma $ located at the entangle-disentangle boundary
is given, entangled states to which $\sigma $ is closet can be obtained
analytically \cite{Ishizaka}. However, the inverse process of obtaining the
optimal state $\sigma $ for a given entangled state is generally not
available. We in this paper deal with a special kind of entangled state. It
is known that by local filtering any two qubit state can be transformed to
either Bell diagonal state or the state of the form \cite{Verstraete}
\begin{equation}
\rho =\frac 12\left[
\begin{array}{llll}
a+c & 0 & 0 & d \\
0 & 0 & 0 & 0 \\
0 & 0 & b-c & 0 \\
d & 0 & 0 & a-b
\end{array}
\right] .  \label{wave1}
\end{equation}
The entanglement property of Bell diagonal state is well known. We in this
paper will investigate the relative entropy of entanglement of this second
local filtering result state. More generally, we consider the following
state which contains the above state as a special case:
\begin{equation}
\rho =\left[
\begin{array}{llll}
\frac 12(\lambda _{+}+\lambda _{-}\cos \phi ) & 0 & 0 & \frac 12\lambda
_{-}\sin \phi e^{-i\eta } \\
0 & \lambda _1 & 0 & 0 \\
0 & 0 & \lambda _2 & 0 \\
\frac 12\lambda _{-}\sin \phi e^{i\eta } & 0 & 0 & \frac 12(\lambda
_{+}-\lambda _{-}\cos \phi )
\end{array}
\right] .  \label{wave2}
\end{equation}
The other two eigenvalues of $\rho $ are $\lambda _0$ and $\lambda _3$, with
$\lambda _{+}=\lambda _0+\lambda _3$ and $\lambda _{-}=\lambda _0-\lambda _3$%
. This state appears at the issue of continuous variable to qubit mapping
\cite{Mista} \cite{Paternostro}.

\section{The relative entropy of entanglement}

Suppose the relative entropy of entanglement of state $\rho $ be reached by
a separable state $\sigma ^{*}$ which has been called as the closet
disentangled state, so that the quantum relative entropy of $\rho $ with
respect to $\sigma ^{*}$ is the smallest. We will find $\sigma ^{*}$ in two
steps, firstly suppose $\sigma ^{*}$ have the same matrix structure as $\rho
$,
\begin{equation}
\sigma ^{*}=\left[
\begin{array}{llll}
\frac 12(\chi _{+}+\chi _{-}\cos \theta ) & 0 & 0 & \frac 12\chi _{-}\sin
\theta e^{-i\varphi } \\
0 & \chi _1 & 0 & 0 \\
0 & 0 & \chi _2 & 0 \\
\frac 12\chi _{-}\sin \theta e^{i\varphi } & 0 & 0 & \frac 12(\chi _{+}-\chi
_{-}\cos \theta )
\end{array}
\right] ,
\end{equation}
the other two eigenvalues of $\sigma ^{*}$ are $\chi _0,\chi _3=\frac
12(\chi _{+}\pm \chi _{-})$. The second step is to prove that such a kind of
$\sigma ^{*}$ is really the state with minimal relative entropy. The full
proof is left to the next section. The separability of two qubit system is
determined by the positivity of the partial transpose of the density matrix.
\begin{eqnarray}
\sigma ^{*PT} &=&\frac 12(\chi _{+}+\chi _{-}\cos \theta )\left|
00\right\rangle \left\langle 00\right| +\frac 12(\chi _{+}-\chi _{-}\cos
\theta )\left| 11\right\rangle \left\langle 11\right| +\chi _1\left|
01\right\rangle \left\langle 01\right| + \\
&&\frac 12\chi _{-}\sin \theta e^{-i\varphi }\left| 01\right\rangle
\left\langle 10\right| +\frac 12\chi _{-}\sin \theta e^{i\varphi }\left|
10\right\rangle \left\langle 01\right| +\chi _2\left| 10\right\rangle
\left\langle 10\right|   \nonumber
\end{eqnarray}
The positivity of $\sigma ^{*PT}$ is determined by the positivity of the
submatrix in the subspace basis $\{\left| 01\right\rangle ,\left|
10\right\rangle \}$, One has $\chi _1\chi _2\geq \left( \frac 12\chi
_{-}\sin \theta \right) ^2$. Furthermore, the state $\sigma ^{*}$ should be
at the border of separable state set (e.g. \cite{Chen}). For a border state,
the separable condition now takes the form of
\begin{equation}
\chi _1\chi _2=\left( \frac 12\chi _{-}\sin \theta \right) ^2.
\end{equation}
This can also be derived from the condition of $C=0,$ where $C$ is the
concurrence of the closet disentangled state $\sigma ^{*}$\cite{Ishizaka}.
Let $\chi _{+}=2A_1\cosh r_1,\chi _{-}=2A_1\sinh r_1,$ $\chi
_1=A_2e^{r_2},\chi _2=A_2e^{-r_2}.$ The separable condition will be $%
A_2=A_1\sinh r_1\sin \theta ,$ The density matrix $\sigma ^{*}$ is in the
form of direct plus of two $2\times 2$ matrices. For a $2\times 2$ matrix
with the form $W=A\cosh rI+A\sinh r\overrightarrow{n}\cdot \overrightarrow{%
\sigma }$, we have $\log W=\log A+r\overrightarrow{n}\cdot \overrightarrow{%
\sigma },$where $\overrightarrow{n}$ is a unit vector and $\sigma _i$ $%
(i=1,2,3)$ are Pauli matrices, Hence
\begin{eqnarray}
-Tr\rho \log \sigma ^{*} &=&-\log A_1-\lambda _{-}r_1[\cos \theta \cos \phi
+\sin \theta \sin \phi \cos (\eta -\varphi )] \\
&&-(\lambda _1+\lambda _2)\log (\sinh r_1\sin \theta )-(\lambda _1-\lambda
_2)r_2.  \nonumber
\end{eqnarray}
Considering the restriction $Tr\sigma ^{*}=1,$ using Lagrangian multiplier
method, we have $f\left( \sigma ^{*}\right) =S\left( \rho \left\| \sigma
^{*}\right. \right) +x$ $\left( Tr\sigma ^{*}-1\right) .$ The
differentiation on $\varphi $ turns out to be $\sin (\eta -\varphi )=0$,
without loss of generality $\varphi =\eta ,$ and from now on $\varphi \ $and
$\eta $ will be omitted. The differentiation on $A_1$ will lead to $x=1.$
The other derivations will be
\begin{eqnarray}
-\lambda _{-}\cos (\phi -\theta )-(\lambda _1+\lambda _2)\coth r_1+2A_1\sinh
r_1+2A_1\cosh r_1\cosh r_2\sin \theta  &=&0, \\
-(\lambda _1-\lambda _2)+2A_1\sinh r_1\sinh r_2\sin \theta  &=&0,
\label{wave3} \\
-\lambda _{-}r_1\sin (\phi -\theta )-(\lambda _1+\lambda _2)\cot \theta
+2A_1\sinh r_1\cosh r_2\cos \theta  &=&0.  \label{wav4}
\end{eqnarray}
Together with $Tr\sigma ^{*}=1$ which now takes
\begin{equation}
2A_1\cosh r_1+2A_1\sinh r_1\sin \theta \cosh r_2=1,
\end{equation}
we can determine $\sigma ^{*}$. For arbitrary $\phi ,$ these equations can
not be analytically solved. Assuming $\phi =\theta ,$ these equations are
solvable. Eq. (\ref{wav4}) has two results, the first is $2A_1\sinh r_1\cosh
r_2\sin \theta $ $=\lambda _1+\lambda _2$, this is a trivial result because
it will lead to the solution $\sigma ^{*}=\rho $ , and the condition $\sin
\phi =2\sqrt{\lambda _1\lambda _2}/\lambda _{-}$ which requires the state $%
\rho $ to be separable; the second result of Eq.(4) is $\cos \theta =0$,
this is a nontrivial solution. Hence when $\phi =\frac \pi 2,$ we have the
solution of $\theta =\frac \pi 2$ and
\begin{eqnarray}
r_2 &=&\log (\sqrt{(\lambda _1-\lambda _2)^2\lambda _{-}^2+4\lambda
_1\lambda _2\left[ 1-(\lambda _1-\lambda _2)^2\right] ^2}-(\lambda
_1-\lambda _2)\lambda _{-}) \\
&&-\log (2\lambda _2(1-\lambda _1+\lambda _2)),  \nonumber \\
r_1 &=&\frac 12\log (1-(\lambda _1-\lambda _2)\tanh \frac{r_2}2)-\frac
12\log (1-(\lambda _1-\lambda _2)\coth \frac{r_2}2), \\
A_1 &=&\frac{\lambda _1-\lambda _2}{2\sinh r_1\sinh r_2}.
\end{eqnarray}
The case of $\phi =\frac \pi 2$ can also be solved by the fact that the
function $-\log x$ is convex which results in
\[
-\log (\left\langle \psi \right| \sigma \left| \psi \right\rangle )\leq
-\left\langle \psi \right| \log \sigma \left| \psi \right\rangle
\]
for any operator $\chi $ and any normalized state $\left| \psi \right\rangle
$ \cite{Vedral2}. Noticing that for $\phi =\theta =\frac \pi 2$ the states $%
\rho $ and $\sigma ^{*}$ have the same eigenbasis. So that
\begin{equation}
E_r\left( \rho \right) =\min_{\chi _i}\sum_i\lambda _i\log (\lambda _i/\chi
_i).
\end{equation}
can be obtained under the constraints $\sum_{i=0}^3\chi _i=1$ and $\chi
_1\chi _2=\frac 14\chi _{-}^2.$The minimization yields analytical solution
of $\chi _i$ as well as $E_r\left( \rho \right) .$ The special case of $%
\lambda _1=0$, that is, the $b=-c$ case of Eq(\ref{wave1}) was already known%
\cite{Vedral3}.

\section{Proof of matrix structure of the separable state}

The most important thing left is to prove that for our special state $\rho $
of Eq.(\ref{wave2}), the separable state $\sigma ^{*}$ has the same matrix
structure as $\rho $. We prove this by the fact that local minimum is also
the global minimum when it is in regard to the relative entropy of
entanglement\cite{Vedral2}. Hence we only need to prove that $\sigma ^{*}$
is the local minimal state. Let $f(x,\sigma ^{*},\sigma )=S(\rho \left\|
(1-x)\sigma ^{*}+x\sigma \right. )$ be the relative entropy of a state
obtained by moving from $\sigma ^{*}$ towards some $\sigma $. The derivative
of $f$ will be \cite{Rehacek}\cite{Vedral2}
\begin{equation}
\frac{\partial f}{\partial x}\left( 0,\sigma ^{*},\sigma \right)
=\int_0^\infty ((\sigma ^{*}+t)^{-1}\rho (\sigma ^{*}+t)^{-1}\delta \sigma
)dt=TrA\delta \sigma ,
\end{equation}
where we denote $(1-x)\sigma ^{*}+x\sigma =\sigma ^{*}-\delta \sigma ,$ and
the operator $A$ has the following matrix elements in the eigenbasis $%
\{\left| \chi _n\right\rangle \}$ of $\sigma ^{*}$:
\begin{equation}
A_{mn}^\chi =\left\langle \chi _m\right| A\left| \chi _n\right\rangle =\frac{%
\log \chi _n-\log \chi _m}{\chi _n-\chi _m}\left\langle \chi _m\right| \rho
\left| \chi _n\right\rangle .
\end{equation}
And when $\chi _m=\chi _n$, the corresponding coefficient should be replaced
with the limit value of $\chi _n^{-1}.$ We have $A_{00}^\chi =\frac 1{2\chi
_0}\left[ \lambda _{+}+\lambda _{-}\cos (\theta -\phi )\right] ,$ $%
A_{11}^\chi =\frac{\lambda _1}{\chi _1},$ $A_{22}^\chi =\frac{\lambda _2}{%
\chi _2},$ $A_{33}^\chi =\frac 1{2\chi _3}\left[ \lambda _{+}-\lambda
_{-}\cos (\theta -\phi )\right] ,$ $A_{03}^\chi =A_{30}^\chi =-\frac{\log
\chi _0-\log \chi _3}{2(\chi _0-\chi _3)}$ $\lambda _{-}\sin (\theta -\phi
). $ And all other $A_{mn}^\chi $ are $0$. Expressing $-Tr\rho \log \sigma
^{*}$ in $\chi _i$ and considering the constraints $\sum_{i=0}^3\chi _i=1$
and $\chi _1\chi _2=\frac 14\chi _{-}^2$, we have
\begin{eqnarray}
-Tr\rho \log \sigma ^{*}-z(\chi _1\chi _2-\frac 14\chi _{-}^2) &=&-\frac
12\left[ \lambda _{+}+\lambda _{-}\cos (\theta -\phi )\right] \log \chi _0 \\
&&-\frac 12\left[ \lambda _{+}-\lambda _{-}\cos (\theta -\phi )\right] \log
\chi _3  \nonumber \\
&&-\lambda _2\log \chi _2-\lambda _1\log (1-\chi _{+}-\chi _2)  \nonumber \\
&&-z((1-\chi _{+}-\chi _2)\chi _2-\frac 14\chi _{-}^2\sin ^2\theta ),
\nonumber
\end{eqnarray}
where $z$ is Lagrangian multiplier. The derivations then will be
\begin{eqnarray}
-\frac 1{2\chi _0}\left[ \lambda _{+}+\lambda _{-}\cos (\theta -\phi
)\right] +\frac{\lambda _1}{1-\chi _{+}-\chi _2}-z(-\chi _2-\frac 12\chi
_{-}\sin ^2\theta ) &=&0, \\
-\frac 1{2\chi _3}\left[ \lambda _{+}-\lambda _{-}\cos (\theta -\phi
)\right] +\frac{\lambda _1}{1-\chi _{+}-\chi _2}-z(-\chi _2+\frac 12\chi
_{-}\sin ^2\theta ) &=&0, \\
\frac{\log \chi _0-\log \chi _3}{2(\chi _0-\chi _3)}\lambda _{-}\sin (\theta
-\phi )-z(-\frac 12\chi _{-}\sin \theta \cos \theta ) &=&0, \\
-\frac{\lambda _2}{\chi _2}+\frac{\lambda _1}{1-\chi _{+}-\chi _2}-z(1-\chi
_{+}-2\chi _2) &=&0.
\end{eqnarray}
From the last equation, we get $z=\frac 1{\chi _1-\chi _2}\left( \frac{%
\lambda _1}{\chi _1}-\frac{\lambda _2}{\chi _2}\right) ,$comparing these
equations with $A_{mn}^\chi $, we get $A_{00}^\chi =\frac{\lambda _1}{\chi _1%
}-z(-\chi _2-\frac 12\chi _{-}\sin ^2\theta ),$ $A_{33}^\chi =\frac{\lambda
_1}{\chi _1}-z(-\chi _2+\frac 12\chi _{-}\sin ^2\theta ),$ $A_{03}^\chi
=\frac 12z\chi _{-}\sin \theta \cos \theta .$ Hence
\begin{equation}
\frac 12(A_{00}^\chi +A_{33}^\chi )=\frac{\lambda _1}{\chi _1}+\frac{\chi _2%
}{\chi _1-\chi _2}\left( \frac{\lambda _1}{\chi _1}-\frac{\lambda _2}{\chi _2%
}\right) =\frac{\lambda _1-\lambda _2}{\chi _1-\chi _2}=1.  \label{wave4}
\end{equation}
The last equality is from Eq.(\ref{wave3}). And $\frac 12(A_{00}^\chi
-A_{33}^\chi )=\frac 12z\chi _{-}\sin ^2\theta .$ So that $\frac
12(A_{00}^\chi -A_{33}^\chi )\cos \theta -A_{03}^\chi \sin \theta =0,$ $%
\frac 12(A_{00}^\chi -A_{33}^\chi )\sin \theta +A_{03}^\chi \cos \theta
=\frac 12z\chi _{-}\sin \theta =z\sqrt{\chi _1\chi _2}$, Hence, in the usual
basis $\left\{ \left| 00\right\rangle ,\left| 01\right\rangle ,\left|
10\right\rangle ,\left| 11\right\rangle \right\} ,$ the operator $A\ $will be

\begin{equation}
A=\left[
\begin{array}{llll}
1 & 0 & 0 & D \\
0 & B & 0 & 0 \\
0 & 0 & C & 0 \\
D & 0 & 0 & 1
\end{array}
\right]
\end{equation}
with $B=\frac{\lambda _1}{\chi _1},$ $C=\frac{\lambda _2}{\chi _2},$ $D=%
\frac{\sqrt{\chi _1\chi _2}}{\chi _1-\chi _2}\left( \frac{\lambda _1}{\chi _1%
}-\frac{\lambda _2}{\chi _2}\right) .$ We should prove that for any
separable state $\chi $,
\begin{equation}
\frac{\partial f}{\partial x}\left( 0,\sigma ^{*},\sigma \right) =TrA\delta
\sigma =TrA(\sigma ^{*}-\sigma )\geq 0.
\end{equation}
where $TrA\sigma ^{*}=1$\cite{Rehacek}, But any $\sigma \in \mathcal{D}$ can
be written in the form of $\sigma =\sum_ip_i\left| \alpha ^i\beta
^i\right\rangle \left\langle \alpha ^i\beta ^i\right| $ and so $\frac{%
\partial f}{\partial x}\left( 0,\sigma ^{*},\sigma \right) =\sum_ip_i\frac{%
\partial f}{\partial x}\left( 0,\sigma ^{*},\left| \alpha ^i\beta
^i\right\rangle \left\langle \alpha ^i\beta ^i\right| \right) $, The problem
is reduced to prove that for any normalized pure state $\left| \alpha \beta
\right\rangle \left\langle \alpha \beta \right| ,$%
\begin{equation}
\left\langle \alpha \beta \right| A\left| \alpha \beta \right\rangle \leq 1.
\label{wave5}
\end{equation}
Let $\left| \alpha \right\rangle =\cos \frac{\theta _1}2\left|
0\right\rangle +\sin \frac{\theta _1}2\exp (i\varphi _1)\left|
1\right\rangle $, $\left| \beta \right\rangle =\cos \frac{\theta _2}2\left|
0\right\rangle +\sin \frac{\theta _2}2\exp (i\varphi _2)\left|
1\right\rangle $ be the most general pure states. The maximum of $%
\left\langle \alpha \beta \right| A\left| \alpha \beta \right\rangle $ over
all $\left| \alpha \beta \right\rangle $ then will be
\begin{eqnarray}
\max_{\alpha ,\beta }\left\langle \alpha \beta \right| A\left| \alpha \beta
\right\rangle  &=&\frac 14(2+B+C+2D+\frac{(B-C)^2}{2-B-C+2D}) \\
&=&1+\frac{D^2-(B-1)(C-1)}{4(2-B-C+2D)}.  \nonumber
\end{eqnarray}
While $D^2-(B-1)(C-1)=\frac{\chi _1\chi _2}{\left( \chi _1-\chi _2\right) ^2}%
\left( \frac{\lambda _1}{\chi _1}-\frac{\lambda _2}{\chi _2}\right)
^2-\left( 1-\frac{\lambda _1}{\chi _1}\right) \left( 1-\frac{\lambda _2}{%
\chi _2}\right) =0.$ The last equality is due to Eq.(\ref{wave4}). So that $%
\max_{\alpha ,\beta }\left\langle \alpha \beta \right| A\left| \alpha \beta
\right\rangle =1.$ And inequality (\ref{wave5}) is proved. We complete the
proof that $\sigma ^{*}$ is the separable state having the smallest quantum
relative entropy of $\rho $ with respect to.

\section{Conclusions and Discussions}

The main result of this paper is that for two qubit state with six none zero
density matrix elements ( besides diagonal elements, the other two none zero
elements are $\left| 00\right\rangle \left\langle 11\right| $ and $\left|
11\right\rangle \left\langle 00\right| $ items) the separable state reaches
the relative entropy of entanglement has the same density matrix structure
as the original entangled state. The proof is strict. The direct application
is the relative entropy of entanglement of two qubit local filtering result
state and more generally two qubit state converted from quantum continuous
variable\cite{Mista} \cite{Paternostro}. It is anticipated that the proof
can be applied to a more general block diagonal state, that is, a state with
eight none zero elements (the new none zero elements are added at $\left|
01\right\rangle \left\langle 10\right| $ and $\left| 10\right\rangle
\left\langle 01\right| $ items ).

\section{Acknowledgement}

Funding by the National Natural Science Foundation of China (under Grant No.
10347119) and Zhejiang Province Natural Science Foundation (Fund for
Talented Professionals, under Grant No. R104265) are gratefully acknowledged.

\end{document}